# WEEPING AND GNASHING OF TEETH:
# TEACHING DEEP LEARNING IN IMAGE AND VIDEO PROCESSING CLASSES

*Al Bovik*

Director, Laboratory for Image and Video Engineering
The University of Texas at Austin

## ABSTRACT

In this rather informal paper and talk I will discuss my own experiences, feelings and evolution as an Image Processing and Digital Video educator trying to navigate the Deep Learning revolution. I will discuss my own ups and downs of trying to deal with extremely rapid technological changes, and how I have reacted to, and dealt with consequent dramatic changes in the relevance of the topics I've taught for three decades. I have arranged the discussion in terms of the stages, over time, of my progression dealing with these sea changes.

## 1. BLISSFUL IGNORANCE

In the year 2012, I had been giving undergrad classroom lectures on Digital Image and Video Processing for more than two decades. My prepared course materials covered the essential topics of image formation, sampling and Fourier theory, wavelets, linear and nonlinear filtering, image coding, and feature detection using LoG and DoG, SIFT and SURF, using them for picture quality prediction, texture analysis, edge, boundary, and line detection, visual search, 3D stereo ranging, motion, and optical flow. I updated my topics as technology progressed, but changes large enough to affect my courseware were easy to keep up with.

Along with hundreds of PowerPoint slides, I had developed dozens of image processing programs with which I could demo the effects of varying algorithm parameters on visual outcomes in real time [1]. I taught perceptual theory, and showed dozens of visual illusions (a hobby of mine) which helped students understand how we see and even affect certain algorithms. Years later, some students remembered these illusions when they chanced upon other ones, which they passed to me, often to join the others in class. I used my own book as a supplement, always had good attendance and great fun (the topic sells itself!), received excellent course ratings, and hardly had to prepare for each lecture. I even wrote about my educational methods for image processing in IEEE journals [2], [3]. Overall, teaching was a blissful experience, and as I thought, would be for decades longer.

## 2. DENIAL

However, also in 2012, the famous "AlexNet/ImageNet" paper was published [4]. Needless to say, a "bomb" had dropped on the machine learning and computer vision communities. Overnight, the attentions of workers in those fields turned to deep models of anything with large data, and spectacular results were obtained on natural language processing and most problems in computer vision. Chatting about this with an ML colleague, I observed "They sure are getting cited a lot, aren't they? To which he replied, "Um, they are the *only* papers being cited!" (with a nervous smile)

Well, this wasn't the case in the field of image processing, and I saw little need to worry much. After all, those researchers were primarily interested in classifying and identifying objects and words, whereas we dealt with pictures intended for people to look at, not robots! Why hop on the neural net bandwagon? I was old enough to have already seen three or four machine learning waves (Perceptrons, Artificial Neural Nets, ConvNets, and Support Vector Machines), each accompanied by considerable hype and some degree of letdown, at least in terms of real-world applications!

Yet it was at least worth chatting about the "deep learning wave" with my class, usually accompanied by the observations just made, and noting that "Well, it is perhaps no surprise that incredibly powerful GPUs crunching to optimize a model having 40 Million free parameters might do pretty well!" and "Don't forget what von Neuman said: 'With four parameters I can fit an elephant, and with five I can make him wiggle his trunk!'" Anyway, I felt no need to start lecturing on these cumbersome algorithms that took days, weeks, or even months to train. Interesting, yes, but not very practical! As a friend said, who is a very notable neurobiologist and bioengineer: "Deep Learning? Bah! Glorified curve fitting!" Well, as my students would say, he was *not wrong,* but …

My work on the research side rather reinforced these rather negative viewpoints. With an interest in its possibilities, Deepti Ghadiyaram and I published one of the earliest papers, if not the first, on picture quality prediction using deep (belief) nets [5]. Unfortunately, when trained and tested on the new LIVE Challenge picture quality database, which was the largest database then available [6] (but see [7]), the deep models could not attain the performance of simple natural scene-based models like BRISQUE [8] and FRIQUEE [9]. Rigorous tests on other researchers' deep models produced similar results [10].

It was becoming clear that these deep models needed much more data than was available in the field of perceptual image processing, where careful psychophysics experiments were required. Unlike human participation in crowdsourced picture labeling experiments like ImageNet [11], where each human label might need only 0.5 – 1.0 sec. to apply, human quality judgments on pictures generally required 10-20x that amount to time for a subject to feel comfortable in making their assessments on a Likert scale [6]. In other words, I was able to hide behind a Great Wall of Non-Existent Perceptual Data, even if other areas of image processing were being infiltrated, which they weren't yet.

After all, popular picture and video quality algorithms still ruled the world! JPEG and MPEG compressed most moving bits, a figure that now approaches 75% of all Internet data. Our own algorithms for picture and video quality were being used throughout industry and Silicon Valley was keeping my graduate students quite busy. Only Deep Learning papers were cited? Ours were doing pretty well. What, Me Worry? [12].

Back in the classroom, students still loved learning about image processing, and I kept updating it with things like Haar cascades for face detection, natural scene statistics models and no-reference picture quality, and more. Deep networks? Well, they hadn't impacted image processing much yet! Generative Adversarial Networks (GANs) and all those synthetic people pictures? Hmmm … seemed like learned database interpolation to me!

Anyway, the field had so expanded in terms of both material and student interest, that in 2014 I created a new graduate class on Digital Video, replete with hundreds of playable video examples showing every aspect of the various parameters, loads of spatio-temporal perceptual theory, and a high level of math rigor for a video class. However, no deep learning at all. Why would I? There wasn't any work to speak of on deep video analysis anyway. What, train a network for 6 months? On what dataset? What kind of hardware would this run on in the real work? Will there be GPUs on the same SOC as MPEG video decoders? Will they (industry) stack the chips? Surely not!

## 3. REALIZATION

Anyway, even if an old dog cannot easily learn new tricks, s/he can at least still watch the puppies play and master them. Researchers more perspicacious than I, or at least readier to try "the new thing," began to apply Deep Learning models to practical problems of regression on images. The results they obtained were provocative and often state-of-the-art (SOTA). A good example is the simple residual based denoiser of my colleague Lei Zhang's group [13]: a simple network and data handling process leading to exceptional results. Plenty of other researchers were getting great results on image processing problems using varieties of deep nets. True, they didn't operate in my space of perception-based analysis, where I had my Great Wall of Non-Data, but it was obvious that things were changing.

On the research side, we were beginning to deploy more sophisticated machine learning models, and especially methods of Transfer Learning, exploiting the generalizability of deep networks, perhaps their most amazing property, and in the news, there were stories of high school "Kagglers" solving all kinds of interesting problems using ultra-accessible new libraries devoted to "Making Deep Learning Easy."

Most disturbing of all were my students' class projects. In all my classes I assign a semester-long Class Project, with a possible reward of Best Project culminating in not having to take the Final Exam! The projects that are produced to compete for an automatic 'A' on the final are always amazing, and are helped along by the class demos at the end of the semester, where the students voted on the winners, rather than a finicky professor. What bothered me a bit was that the projects increasingly used Deep Learning, none of which I taught in the class! The solutions were better, and the problems solved were more complex (e.g., gesture recognition, image compression, style transfer, and much more). Even as my own hairs were getting grayer, I began to feel that my teaching of image processing, an important part of my professional identity that I took very seriously, was in danger of becoming obsolete. Soon, I figured the students would be whispering "Old Boomer, he's still living in the past! I'd take the computer vision class over in CS instead."

## 4. DISPAIR

Sure enough, by 2016, I was beginning to notice the effects on my teaching program. Attendance was noticeably dropping in my classes. At first, I attributed this to just part of the irrational rush into Data Science, when in the Fall attendance in my image processing class fell by 30% from the previous years. Even worse, in the Spring of 2017, a promising enrollment of 25 students in my advanced grad class Digital Video fell precipitously after the first week. A couple of students asked me after class, "I didn't see Deep Learning on the syllabus, will you be covering that?" I was a bit dismayed to tell them "Well, not this time. It hasn't really started to impact video yet." Which, truth be told, was not quite true anymore. In any case, a week later enrollment was at 14, while the new Large-Scale Optimization class being offered had to turn away students after reaching the room limit of 100. OK, Boomer.

It suddenly felt a bit odd trudging through the rather heavy math I offered throughout the early parts of the class, once a highlight of the material: developing high-dimensional transforms and the theory of space-time distributions. But how can motion in videos be understood without covering singularity theory?

But of course, parallel to my own woes, things were changing everywhere. One-time High Impact champion *IEEE Transactions on Information Theory* was beginning its descent to levels just above the *IEEE Journal of Oceanic Engineering,* where it currently languishes. I cast no aspersion on either, and have published in both. I would wince when I would read papers pronouncing new heights in deep network performance on problems old and new, some unimaginable, where the authors would inevitably compare their results with "the *handcrafted* method of so and so." Not quite pejorative, but getting there. How brilliant to instead shovel data into a computing engine! What of all those years of actually thinking, of dissecting the little knowledge we had of brain function, developing and deploying models of neural function to create amazing and useful picture processing algorithms? I began to feel like an old wood carver slowly chipping away outside a great and shining high tech city, creating interesting little doo-dads, watching each passerby, hoping to catch someone's eye.

The less indelicate of these Computer Science authors would instead refer to anything using features, no matter if derived from neurophysiological measurements, statistical physics, or mathematical deduction as …. "engineered." Thanks for that!

This was mainly a concern in the classroom, of course. My course materials were clearly becoming obsolete at an astonishing pace. On the research side, my students were adaptive and adopting all kinds of machine learning methods into their creative efforts. As always, I learn more from them, than they from I. Something had to change, however, in my teaching of image and video processing, and soon.

### 5. ACCEPTANCE

My method of instruction tends towards the exceedingly prepared. Each semester, I charge through about 1000 PowerPoint slides at a rate of 40/day, showing live-action visual examples throughout. While I was once comfortable chalking away clouds of dust teaching DSP, it isn't practical when students need to see the results, and when the volume of material has become immense given the cross-disciplinary nature of the field and the explosion of applications. There is huge ground to cover. In any case, creating substantial modifications in a massively prepared class like mine is like steering a battleship. Nevertheless, I steeled myself to do just that, one class after the other, beginning with the undergraduate image processing class. To enable this, I took a paid faculty leave in the Spring of 2018, so that I could focus on recreating "Digital Image Processing" in the Deep Learning Age. I did not suspect that I would need to do it again the following Spring, to be able to accomplish the same thing in the Digital Video class.

### 6. ACTION

Once resolved, I eyeballed my notes and realized that the task ahead would require hundreds of hours of work. I wanted the class to be special and complete, as it had once been, and in particular, I wanted it to be immersively visual. We are all visual creatures, with about 50% of brain function implicated in sight, but I think I am far more so, so that is my preferred way of teaching as well. Wherever I travel, my first stop is at the local gallery, and visits to London, Paris, and Florence are tireless hikes to see every great work of art. For me, a week without visiting the cinema leads to withdrawal symptoms. Anyway, the task ahead seemed daunting.

For the image processing class, I wanted to immerse the students from beginning to end in both classical image processing as well as machine learning. How to do both? I decided that, since many students would enter class without *any* exposure to machine learning (as with many schools, we are unprepared still to handle this cresting wave), I would have to start at the beginning. In this way, I could also make my own command of the subject matter completer and more authoritative.

So, I created hundreds of slides on machine learning, beginning with the Perceptron. It took many months, hundreds of hours of creating colorful slides, diagrams, and visual examples. I covered Multi-Layer Perceptrons (MLPs) and backprop, radial basis functions, and support vector machines, arriving at ConvNets, revealing the amazing VGG-16 [14] which we would dissect in detail, a wonderful exemplar of many of the principles of deep learning, and still popular today as a "deep feature model." Then on to ResNet [15], the biggest advance since AlexNet and ImageNet, autoencoders, and the wonders of transfer learning. I was determined to find and give instruction on the most interesting and important applications of Deep Learning in denoising, image compression, picture quality prediction, recognition, and computational stereopsis, while retaining as much fundamental "classic" material as I could.

The problem was, what to retain? So much material had to be, simply, eliminated, like the vacuum tube. After much contemplation, things began to be surgically excised, like mathematical morphology (except on binary pictures), anisotropic diffusion, order statistic filters, and AM-FM image models. I chewed my lip over edge detection, and finally retained half of it, if only to exemplify the utility of the image gradient, and to explain connections between retinal processing (DoG, LoG) and image processing applications, like SSIM [16], BRISQUE, SIFT [17], and SURF [18]. Finally, this long process was over, just as the Fall 2019 semester was commencing.

I advertised the class as Digital Image Processing in a flyer, with a list the topics covered, new and old, emblazoned by "WITH DEEP LEARNING!" superimposed. Before long, 100 students had signed onto the class, and nearly all (90 of them) stayed the course, and I had more fun than I could have imagined, both reveling in the magical results of deep models, and as I discovered, combining my teachings on old and new ideas to create a richer learning experience. After all concepts like ResNet can be understood not only from the data science perspective, (allowing very deep models to be built with better convergence, and less overfitting or vanishing gradients) but this is hardly a surprise to neuroscientists. After all, the neurovisual system has evolved not only to optimal filters, like Gabor functions similar to the early layers of an ImageNet-trained deep net, but also to exploit the considerable redundancies in visual data via processes of normalization similar to residual coding. From that perspective, it all dates back to Barlow's Efficient Coding Hypothesis [19].

In any case, the course was successful, the student projects were more amazing than ever, the feedback I received was uniformly positive, and all in all, it was enormously satisfying to find myself once again teaching things that were SOTA and relevant to the moment.

Of course, I still had my graduate course to consider, and so, now emboldened, I repeated the process, once again wielding the scimitar, but with even more energy, as great swathes of mathematics and proofs, sampling theory, wavelet theory and filter theory were simply axed. To preserve my dignity, I kept much of this material, linking it to the Lectures, where appropriate, in case there were any students remaining who appreciated the analysis side of Digital Video. Yes, but wavelet theory? Gone? Well yes … after all, neurophysiological systems, like deep nets it seems, seek efficient processing through massively overcomplete representations. No need to discuss Perfect Reconstruction.

After all, in addition to increasing the sophistication and timeliness of the deep learning material to include modern innovations like DenseNet [20] and ResNext [21], I needed to also lecture on global-to-local networks like Faster R-CNNs [22], space-time networks like Flownet [23], GANs, and how to use these in advanced applications in motion estimation, video compression, video quality prediction, and more. In fact, it is mid-semester into this class as I write this little paper, and I have just finished placing the finishing touches on those notes, and just in time, as the class lectures are catching up to the notes. I am having fun again, I have 35 students in the class, and I am very much looking forward to the class projects presentations and demonstrations on Digital Video, most involving Deep Learning models, in the final 2-3 days of the semester. I am sure the students will surprise and amaze me as they always do.

## 7. HOW I LEARNED TO STOP WORRYING AND LOVE THE BOMB

I have never actually been interested in the topic of neural networks. Other than, long ago, understanding the properties of MLPs, backpropagation, and the universal approximation theorems, I have largely ignored them. Sure, I think everyone understood their eventual potential, and that indeed, one day they would become the key to artificial intelligence. After all, massive connectivity, and all that, right? Still, they are nothing more than optimization machines with zillions of parameters, why shouldn't they be able to learn just about any dataset? Indeed, they are still boring (to me) in that way, since I am generally only excited about things related to *understanding* visual processing and visual perception, and a black box that generates high F1 scores is not that.

However, Deep Learning engines are definitely *not* boring in that they can be explained as achieving the optimizations that we have long taught of multiscale processing, sparsity, space-frequency localization, linear and nonlinear models of feature abstraction, and adaptation to the natural statistics of the visual world. Even more so, they are unlimited in their applications, and in their ability to realize concepts that we have tried to solve for many years, but have been only poorly implemented in "the old ways." In the future, I think we will see Deep Networks being used to control the efficiency of most of the data being transported over the Internet, nearly all of which will be pictures and videos. It will be exciting to be part of that for the few decades I have left conducting research and teaching, and seeing the kinds of applications I have imagined for many years, dating long before most people knew what a digital picture was, realized at the largest scales. It will be even more satisfying to lecture and teach these topics, since this is why I come to the University, far more so than to conduct research, despite all the energy I put into that.

## 8. NIRVANA

I think that a lot of other professors of image and video processing, as well as of computer vision, must have had similar experiences as I have. I haven't been able to fully express how much the Deep Learning revolution has affected me, but I can say that it has substantially affected my professional life to a greater degree than anything since my assistant professor years 35 years ago. It's also given me a sense of accomplishment to "catch up to" this wave in the classroom. I hope to hear from others their own experiences and ways of dealing with these major changes. As always, all of my course materials are freely available for others to use. You can preview the course notes (in PDF form without links, programs or visual illusions) on the LIVE website [24], and I am happy to share the PowerPoints in their entirety on request. After all, teaching is blissful again!